# Supercontinuum-driven broadband Fourier-transform photothermal microspectroscopy

Senyu Wang [1,2], Zicong Xu [1,3], Kazuki Hashimoto [1,3,4], Jianfeng Li [2,*], Takuro Ideguchi [1,3,4,**]

1 Institute for Photon Science and Technology, The University of Tokyo, Tokyo, 113-0033, Japan

2 School of Optoelectronic Science and Engineering, University of Electronic Science and Technology of China, 610000, Chengdu, China

3 Department of Advanced Materials Science, Graduate School of Frontier Sciences, The University of Tokyo, Chiba, 277-8561, Japan

4 RIKEN Center for Advanced Photonics, RIKEN, Saitama, 351-0198, Japan

* lijianfeng@uestc.edu.cn

** ideguchi@g.ecc.u-tokyo.ac.jp

## Abstract

Chemical characterization of microscopic samples often requires resolving multiple, widely separated vibrational absorption bands, including their positions and spectral shapes. Photothermal microspectroscopy provides a route to localized vibrational measurements by detecting absorption-induced thermal responses with a separate optical probe. However, rapid acquisition of broadband photothermal spectra remains challenging when spectral information is collected through sequential wavelength tuning. Here, we develop a supercontinuum (SC)-driven Fourier-transform photothermal (FT-PT) microspectroscopy platform that combines broadband near-infrared excitation with rapid-scan interferometric acquisition for localized vibrational measurements. In this approach, an interferometrically modulated SC pump induces wavelength-dependent photothermal responses in the sample, which are detected through a co-propagating visible probe beam and reconstructed by Fourier transformation. Using a 1.65–2.35 μm spectral window, we demonstrate broadband photothermal spectra of aqueous samples with acquisition rates of approximately one spectrum per second. The system simultaneously resolves multiple molecular absorption features in a water–dimethyl sulfoxide mixture and reproduces the near-infrared absorption characteristics of liquid water with high spectral fidelity. Extending beyond homogeneous liquids, a dark-field detection configuration enables broadband photothermal spectroscopy of an individual 6 μm polystyrene particle in an aqueous environment. This SC-based FT-PT microspectroscopy approach provides a rapid and broadband platform for localized vibrational characterization of liquid and microscopic samples, enabling chemical analysis with microscale spatial selectivity.

## Introduction

Vibrational spectroscopy identifies molecules and their environments through absorption bands whose positions, shapes, and relative amplitudes jointly report on chemical composition, intermolecular interactions, and local physical state. Diagnostic bands may lie hundreds to more than a thousand wavenumbers apart, making it valuable to compare them in one acquisition on a common spectral axis. Fourier-transform infrared (FTIR) spectroscopy provides broadband, multiplexed acquisition [1, 2], while dual-comb and time-stretch spectroscopies have increased the speed of broadband measurements [3–6]. These methods read out light transmitted or reflected by the sample, so absorption is inferred from the intensity change on the source spectral background. For a small sampling volume and short effective absorption path, that fractional change can be especially difficult to resolve.

Photothermal (PT) spectroscopy addresses this difficulty by detecting the local thermal response to absorbed infrared light, rather than resolving a small attenuation of the infrared beam [7–9]. Non-radiative relaxation converts absorbed optical energy into heat, producing a local temperature change and thus a refractive-index perturbation. PT spectroscopy and microscopy detect this perturbation with a separate visible probe; changes in the probe's phase or propagation are converted into a measurable intensity modulation [10–17]. Only absorbed infrared wavelengths generate a PT signal; unabsorbed light passes through the sample without contributing. The resulting PT spectrum is therefore free of the large infrared spectral background. The infrared pump selects the molecular vibration, while the overlap of pump and probe localizes the readout; point-scanning and wide-field arrangements provide different ways to map this response [10–17]. Mid-infrared PT microscopy accesses strong fundamental vibrations [10–12, 16, 17]. At shorter infrared wavelengths, weaker overtone and combination bands provide access to selected windows useful for measurements in water-rich or scattering samples: the C–H second overtone near 1.2 μm [18], the C–H first overtone near 1.7 μm [13], and C–H combination bands at 2.0–2.5 μm [19]. Many wavelength-resolved PT microscopes use tunable excitation, including quantum cascade lasers (QCL) and optical parametric oscillators [10–13, 16, 18]. Their spectra are collected by sequentially scanning the wavelength, with coverage set by each source's tuning range. Comparing widely separated bands can therefore require extended sweeps or multiple sources.

Multiplexed PT methods address this sequential-acquisition constraint. Dual-comb PT spectroscopy has been demonstrated for gas detection [20], applied to microscopic sample imaging [21], and combined with fluorescence-detected PT imaging for mid-infrared spectromicroscopy [22]. Fourier-transform modulation of the excitation has also enabled PT gas spectroscopy using a QCL frequency comb [23] and broadband optical photothermal infrared spectroscopy and imaging using synchrotron radiation [24]. The cited comb-based measurements target individual bands or limited spectral regions: the acetylene $\nu_1+\nu_3$ band over about 1 THz [20], an approximately 70 cm$^{-1}$ window of nitrous oxide [23], and the fundamental C–H stretching region near 3.4 μm in microscopic samples [21]. Synchrotron radiation spans a much broader range, but its reported PT implementation requires a facility source and uses step-scan interferometry with lock-in detection at each mirror position [24]. Rapid acquisition of several widely separated bands from a condensed-phase specimen with a table-top source thus remains a challenge.

Here we report supercontinuum (SC)-driven broadband Fourier-transform photothermal (FT-PT) microspectroscopy in the near infrared. A fiber SC pump spans widely separated near-infrared overtone and combination bands with the spatial coherence needed for tight focusing. The source delivers nanosecond, microjoule-level pulses at 60 kHz, which produce detectable PT signal even in these relatively weak absorption bands. Its 1.65–2.35 μm output (approximately 4250–6060 cm$^{-1}$) is modulated by a continuously scanned Michelson interferometer (MI) and focused with a 515-nm continuous-wave probe through a reflective objective. A silicon photodetector records the resulting PT modulation of the probe. For a 50:50 (v/v) dimethyl sulfoxide (DMSO)–water mixture at 28 mW average pump power, individual pulses produce a modulation depth approaching 10%. This response permits direct digitization of PT interferograms during continuous mirror scanning, without a separate lock-in demodulation stage, at approximately one interferogram per second and a resolution of 10 cm$^{-1}$. The spectrum is acquired without tuning the pump and reconstructed on a common wavenumber axis referenced to a helium–neon (He–Ne) laser. In the DMSO–water mixture, a weak C–H first-overtone feature near 1.7 μm, the O–H combination band near 1.93 μm, and C–H combination bands between 2.2 and 2.35 μm are recovered within the same broadband measurement. With dark-field detection, we also record characteristic absorption features from a single 6-μm polystyrene (PS) bead in heavy water ($D_2O$), extending the measurement from a homogeneous liquid to a discrete microscopic absorber.

## Results

### Operating principle and experimental implementation of SC-driven FT-PT microspectroscopy

Figure 1 summarizes how the SC-driven FT-PT microspectroscopy system converts broadband infrared absorption into a probe-beam interferogram and then reconstructs the corresponding spectrum. A broadband SC pump is modulated by an MI before being combined collinearly with a visible (VIS) probe. As the interferometer scans, the intensity of each pump spectral component varies with the optical path-length difference (OPD), at a rate determined by its wavenumber. The spectral components absorbed by the sample are labeled $\nu_{\alpha 1}$ to $\nu_{\alpha K}$ in the schematic; their OPD-dependent intensity variations therefore produce a corresponding time-dependent heat load. This heat load generates a refractive-index perturbation that modulates the overlapping VIS probe, producing a PT interferogram. Fourier transformation of the interferogram recovers the corresponding spectral components on the infrared wavenumber axis. The probe therefore carries absorption bands encoded by the broadband pump, allowing them to be acquired in a single scan.

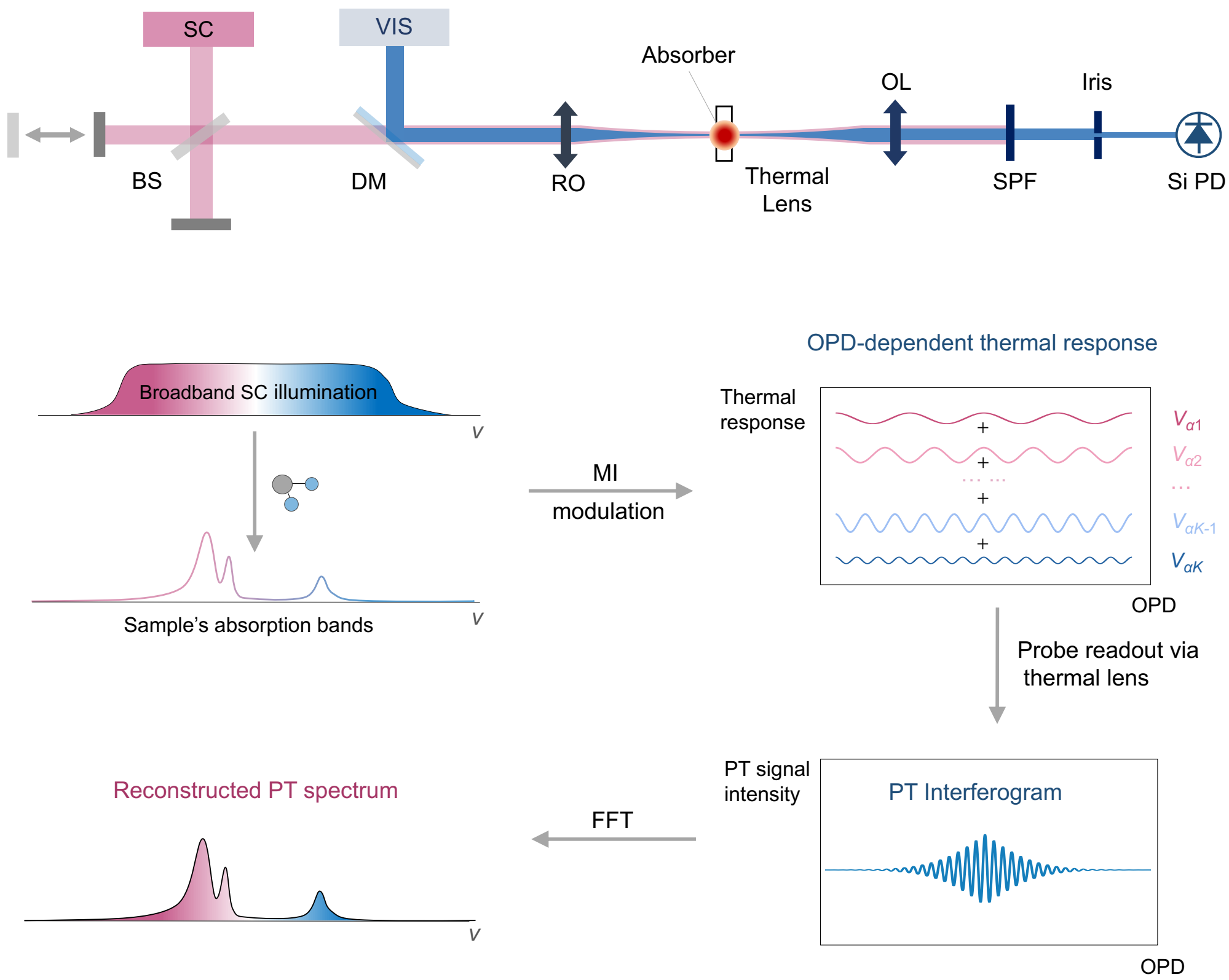

**Fig. 1 Operating principle of SC-driven FT-PT microspectroscopy**. The lower schematic summarizes how sample absorption is encoded by MI modulation as an OPD-dependent thermal response, read out as a probe PT interferogram, and Fourier-transformed into a PT spectrum on the wavenumber axis**.** SC, supercontinuum; VIS, visible; BS, beam splitter; DM, dichroic mirror; RO, reflective objective; OL, objective lens; SPF, short-pass filter; Si PD, silicon photodiode; MI, Michelson interferometer; PT, photothermal; OPD, optical path-length difference; FFT, fast Fourier transform.

The experimental implementation of the SC-driven FT-PT microspectroscopy system is shown in Fig. 2a. The pump is a custom-built, table-top SC source seeded by a commercial 1550-nm laser. The 6-ns seed pulses are delivered at a repetition rate of 60 kHz, amplified in an erbium–ytterbium-doped fiber amplifier, and spectrally broadened through modulation instability and Raman red-shifting in a long single-mode fiber. Fiber-based SC generation can provide bandwidths of several thousand wavenumbers, enabling simultaneous access to widely separated overtone and combination bands [25]. Within the fiber amplifier's power-handling limits, the 6-ns duration provides sufficient peak power for SC generation while retaining microjoule-level pulse energy for weak-band PT excitation. The 60-kHz repetition rate provides a 16.7-μs interval for cooling between pulses, as assessed from the microsecond-scale PT decay below. Detailed source specifications are given in the Methods section.

The fiber-delivered SC pump is collimated by a reflective collimator (RC) and spectrally purified by a tilted 1700-nm long-pass filter before entering the MI. A He–Ne reference laser is coupled into the MI and detected by silicon photodiode

PD1 to track the OPD of the moving mirror. The interferometrically modulated pump is then combined collinearly with a continuous-wave visible probe beam using dichroic mirror DM2, and both beams are focused onto the sample by a reflective Schwarzschild objective (RSO). Both beams filled the NA = 0.8 RSO pupil, giving estimated Airy spot diameters of 3.1 μm for the pump and 0.79 μm for the probe from d = 1.22λ/NA, using wavelengths of 2.0 μm and 515 nm, respectively. An uncoated window before the RSO directs a small fraction of the pump to an extended-range InGaAs photodetector (PD2) for monitoring the output SC spectrum. The filtered SC spectrum corrected for detector responsivity, shown in Fig. 2b, spans approximately 1.65–2.35 μm (4250–6060 $cm^{-1}$). At fixed NA, the estimated focal power density scales as $I_f(\lambda) \propto S_{SC}(\lambda)/\lambda^2$, where $S_{SC}(\lambda)$ is the SC spectral intensity (Fig. 2b). This profile is used for the excitation-profile correction below. After the sample, the transmitted probe is collected by an objective lens (OL); a short-pass filter (SPF) rejects residual infrared light, and an iris transmits the near-axial probe component to silicon photodiode PD3. Because the photothermal response is, to first order, a phase perturbation produced by a transient thermal lens, this spatial filtering converts the angular redistribution of the probe into an intensity modulation. Collecting the entire transmitted probe would reduce this intensity contrast, so the iris diameter was adjusted to maximize the PT signal. The PD3 output was acquired in two modes. For the single-pulse response, the signal was directly amplified in DC coupling mode and digitized at 10 MS/s (route 1). For interferogram acquisition, the signal was DC-blocked, low-pass filtered at 20 kHz, and digitized together with the He–Ne reference at 100 kS/s (route 2).

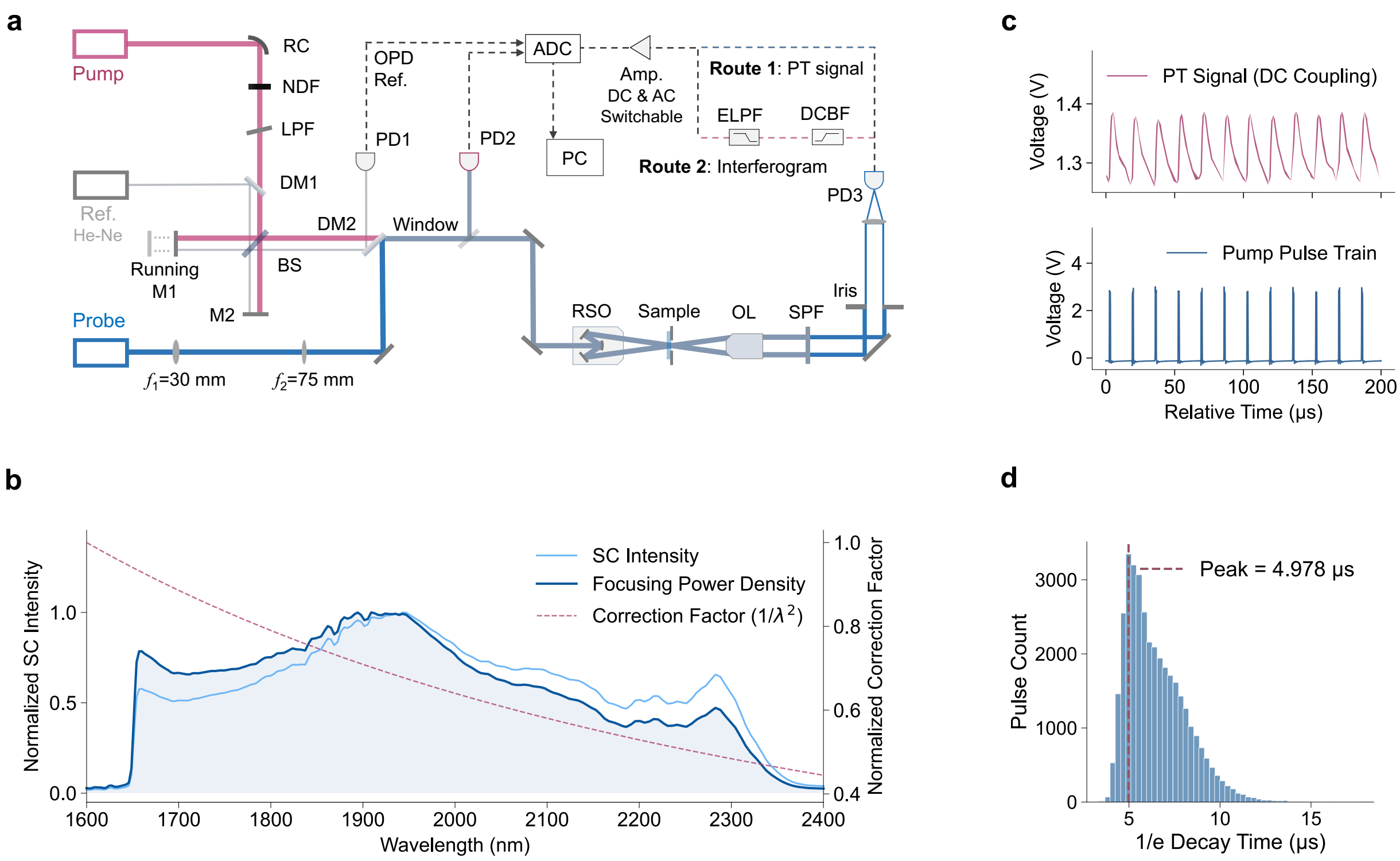

**Fig. 2 Experimental implementation and transient photothermal response of a 50:50 (v/v) DMSO–water mixture.** (a) Optical layout and acquisition routes. RC, reflective collimator; NDF, neutral-density filter; LPF, long-pass filter; DM1–DM2, dichroic mirrors; BS, beam splitter; RSO, reflective Schwarzschild objective; OL, objective lens; SPF, short-pass filter; PD1–PD3, photodiodes; OPD, optical path-length difference; Amp., amplifier; ELPF, electronic low-pass filter; DCBF, DC-blocking filter; ADC, analog-to-digital converter. (b) SC spectral intensity, estimated focal power density, and the $1/\lambda^2$ focusing factor. The curves are normalized for display. (c) Time-resolved probe PT signal (upper) and 60-kHz pump pulse train (lower), recorded through route 1. (d) Distribution of fitted 1/e decay times; the dashed line marks the mean.

### Broadband spectral acquisition and characterization of a DMSO–water mixture

We used a 50:50 (v/v) DMSO–water mixture to assess broadband spectral acquisition and the repeatability of widely separated vibrational bands. The mixture produced a directly measurable PT response to the 60-kHz pump pulse train (Fig. 2c). Each pulse induced a rapid change in the probe signal followed by thermal relaxation. Single-exponential fits yield an effective 1/e decay time distribution peaking at 5 μs, with a mean of 6.5 μs (Fig. 2d; Methods section). For the 50:50 (v/v) mixture, averaging the room-temperature thermal diffusivities of water ($1.43 \times 10^{-7}$ $m^2/s$) and DMSO ($1.05 \times 10^{-7}$ $m^2/s$) [26] gives $D_{mix} \approx 1.24 \times 10^{-7}$ $m^2/s$. Notably, $D_{mix}$ is a simple linear approximation rather than an experimentally determined value for the mixture. With a pump radius r ≈ 1.5 μm, the diffusion estimate $t_D = r^2/(4D_{mix}) \approx 4.5$ μs is comparable to the measured decay time. The 16.7-μs pulse interval at 60 kHz therefore allows substantial cooling between excitations, helping to suppress cumulative sample heating. At an average pump power of 28 mW, corresponding to approximately 0.47 μJ per pulse at the sample, the single-pulse modulation depth, defined as the change in probe-intensity relative to the pump-off probe intensity, approached 10%. The large modulation depth enabled direct digitization of the PT interferogram during continuous scanning. Using acquisition route 2 in Fig. 2a, we recorded successive interferograms (Fig. 3a) together with the He–Ne reference. Resampling at the reference zero crossings established a common OPD grid, and Fourier transformation of the resulting interferograms reconstructed the broadband PT spectra.

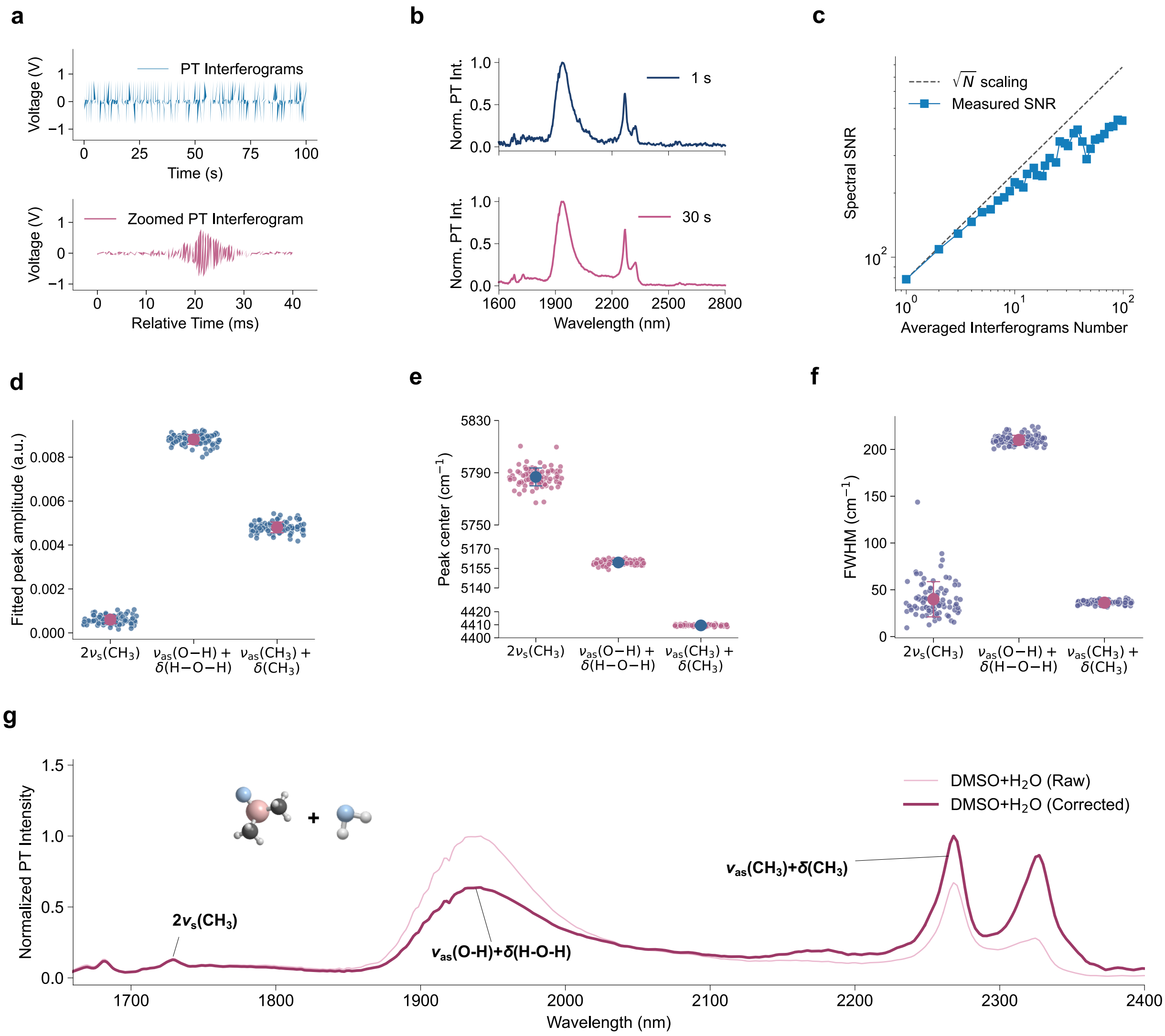

**Fig. 3 Broadband spectral acquisition and repeatability of a 50:50 (v/v) DMSO–water mixture.** (a) Successive PT interferograms over approximately 100 s (upper) and a zoomed centerburst (lower), recorded through route 2 in Fig. 2a. (b) Normalized spectra from 1 s and 30 s acquisitions at 10 $cm^{-1}$

resolution. (c) Spectral signal-to-noise ratio (SNR) versus the number N of coherently averaged interferograms; the dashed line shows $\sqrt{N}$ scaling referenced to the mean SNR of all individual acquisitions. (d–f) Peak amplitude above the local baseline, center wavenumber, and full width at half maximum (FWHM) for the bands assigned to $2\nu_s(CH_3)$, $\nu_{as}$(O-H) + δ(H-O-H), and $\nu_{as}(CH_3) + \delta(CH_3)$. Dots represent valid fits to individual 1-s spectra; error bars indicate mean ± standard deviation (SD). Horizontal jitter separates overlapping dots and has no physical meaning. (g) Normalized raw and excitation-profile-corrected PT spectra after 30-s coherent averaging.

With the moving mirror continuously scanned at 2 mm/s, a symmetric scan around zero OPD, including the turnaround deceleration, yielded approximately one interferogram per second. At this mirror velocity, the 4250–6060 $cm^{-1}$ optical band is encoded at approximately 1.7–2.4 kHz. Direct acquisition captures this interferogram bandwidth without the additional output filtering associated with lock-in demodulation. The 20-kHz acquisition filter lies well above the 1.7–2.4 kHz signal band. Figure 3b compares spectra from a single 1-s acquisition and a 30-s coherent average, with a spectral resolution of 10 $cm^{-1}$. The main absorption bands are already visible in the single-acquisition spectrum, while averaging reduces the fluctuations and improves the visibility of weaker features. Figure 3c quantifies the averaging benefit using the spectral peak amplitude divided by the baseline standard deviation (SD) over 2700–2900 nm. The mean single-acquisition signal-to-noise ratio (SNR) was approximately 77, increasing to approximately 360 after averaging 30 interferograms (approximately 30 s). The SNR increased approximately with $\sqrt{N}$.

To assess spectral repeatability, we characterized individual 1-s spectra by their fitted peak amplitudes, center wavenumbers, and full width at half maximum (FWHM; Fig. 3d–f); the detailed fitting procedure is described in the Methods section. Three bands near 4410, 5160, and 5787 $cm^{-1}$ were selected, corresponding to $\nu_{as}(CH_3) + \delta(CH_3)$, $\nu_{as}$(O-H) + δ(H-O-H), and $2\nu_s(CH_3)$, respectively [27]. The peak-amplitude coefficients of variation were 5.1%, 2.5%, and 30.7%, and the standard deviations of the center wavenumbers were 0.6, 1.6, and 6.8 $cm^{-1}$, respectively. The FWHM estimates were less precise for the weak overtone. The stronger bands near 4410 and 5160 $cm^{-1}$ showed good repeatability, whereas the weak overtone near 5787 $cm^{-1}$ required longer averaging for reliable identification. Figure 3g shows the raw and excitation-profile-corrected PT spectra after 30-s coherent averaging. This correction accounts for the SC spectral envelope and wavelength-dependent focusing shown in Fig. 2b, revealing the weak methyl overtones near 1.7 μm together with the broad water combination band near 1.93 μm and the methyl combination band near 2.3 μm. These bands span approximately 1400 $cm^{-1}$ and are recovered within the same interferometric measurement without tuning the pump wavelength.

**Broadband photothermal microspectroscopy of a 6 μm polystyrene particle**

We measured a single 6 μm PS bead immersed in $D_2O$ using the dark-field geometry shown in Fig. 4a. The central obscuration of the RSO produces an annular directly transmitted probe beam. The collection iris passes probe light scattered by the bead into the central dark region while rejecting most of the unscattered annulus. Local heating perturbs the refractive index and the particle–medium interface, modulating the collected scattered light. This spatial filtering provides the particle-sensitive PT detection [19].

The particle measurement used the same continuous-scan acquisition and He–Ne-referenced reconstruction as the liquid measurements. Figure 4b shows an interferogram acquired in approximately 1 s with the pump and probe focused on the bead. Its centerburst encodes the broadband PT response of the bead–solvent system. The transformed spectrum with a resolution of 10 $cm^{-1}$ is shown in Fig. 4c. This single-acquisition spectrum has an SNR of approximately 29 and contains features from both PS and $D_2O$. The $D_2O$ features indicate that the detected PT spectrum still includes a solvent contribution under dark-field detection. Particle-related features are nevertheless visible in this approximately 1-s acquisition.

To reduce fluctuations and separate the PS spectrum from the solvent contribution, 30 successive interferograms were coherently averaged, corresponding to approximately 30 s of acquisition, and the resulting spectrum was corrected for the excitation profile before the scaled $D_2O$ background was subtracted. The background-subtraction procedure is described in the Methods section. Figure 4d compares the resulting particle spectrum with the reference PS absorption curve calculated from tabulated optical constants [28]. The measurement reproduces the broad feature near 1.68 μm (2*ν*(Ar−C−H)), the closely spaced absorption features around 2.14–2.19 μm (*ν*(Ar−C−H) + δ(phenyl ring)), and the longer-wavelength band beyond 2.3 μm (*ν*($CH_2$) + δ($CH_2$)) [29]. These features are consistent with the C–H overtone and combination absorptions of PS and are recovered together over the broadband measurement window. Several PS-related features are already discernible in the single-acquisition spectrum recorded in approximately 1 s (Fig. 4c). Coherent averaging of 30 interferograms over approximately 30 s reduces fluctuations and makes these features clearer (Fig. 4d). The agreement between the peak positions and profiles in the averaged spectrum and those of the reference supports identification of microscopic PS bead from multiple molecular features. A broad residual feature remains near 2.05 μm, which may arise from incomplete $D_2O$ background subtraction and residual measurement uncertainties. Recovering a broadband molecular spectrum from one selected 6 μm bead in a liquid environment demonstrates FT-PT microspectroscopy beyond homogeneous samples.

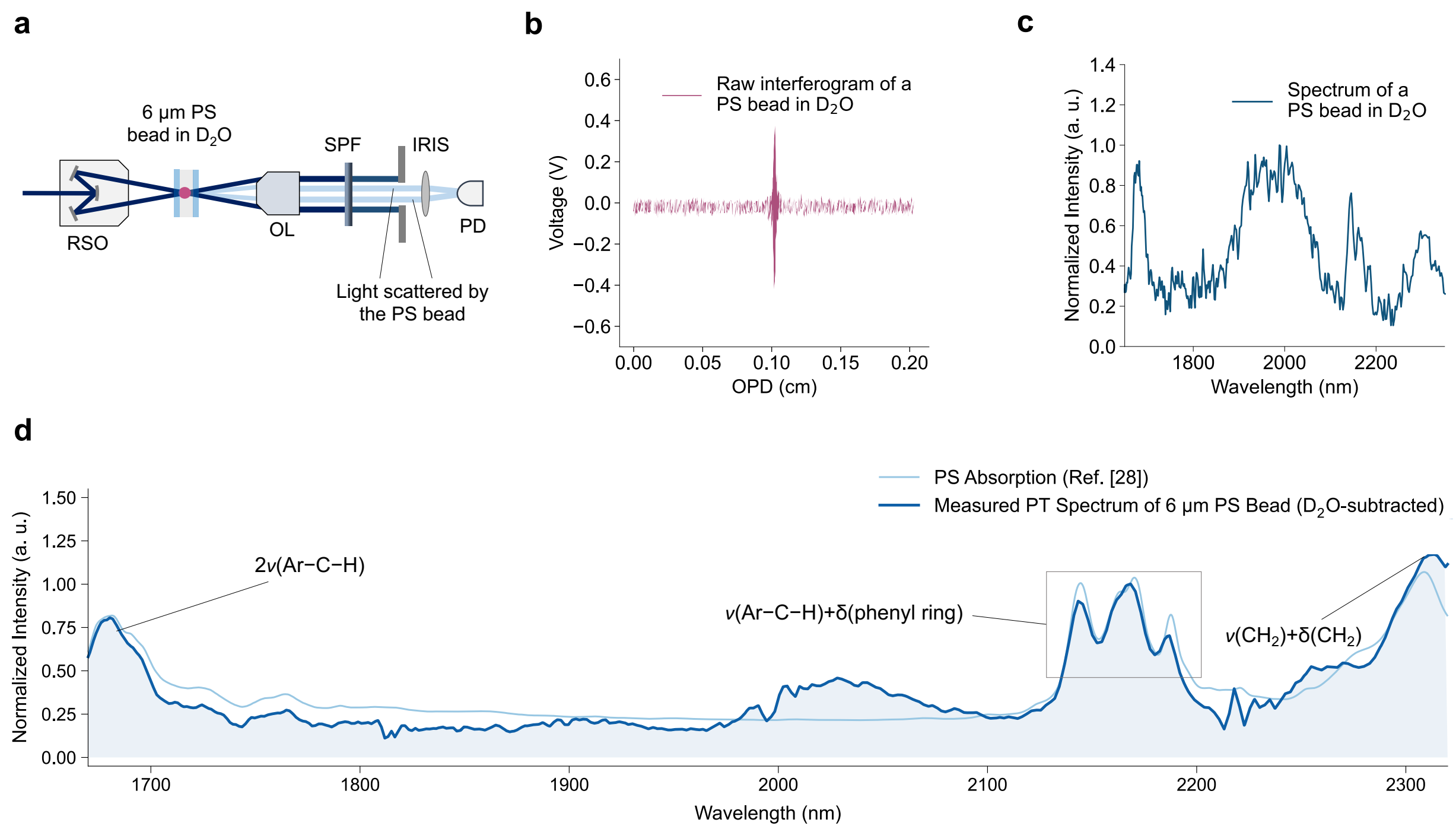

**Fig. 4 Broadband photothermal microspectroscopy of a single 6 μm polystyrene bead in $D_2O$.** (a) Dark-field geometry in which the collection iris passes bead-scattered probe light in the central dark region of the transmitted annular beam while rejecting most unscattered light. RSO, reflective Schwarzschild objective; OL, objective lens; SPF, short-pass filter; PD, photodiode. (b) PT interferogram acquired in approximately 1 s with the focus on the bead. (c) Fourier-transformed spectrum of (b) (SNR ≈ 29). (d) PS spectrum after coherent averaging of 30 interferograms (approximately 30 s), excitation-profile correction, and $D_2O$-background subtraction, compared with the reference PS absorption curve calculated from tabulated optical constants [28].

## Pump-induced spectral perturbations in liquid water

We then examined how excitation power affects SNR and the sample spectrum, using pure water as a model system. At a pump power of 2.4 mW, the normalized PT spectrum agrees well with the commercial FTIR reference in the peak position and overall profile of the absorption band near 1.93 μm (Fig. 5a), supporting its spectral fidelity. Small differences remain on the peak sides, where weaker absorption reduces the PT signal and spectral SNR. The spectral SNR increases approximately linearly with pump power (Fig. 5b) but remains nearly constant as probe power increases (Fig. 5c). These power dependences support a multiplicative noise limit in the current system, consistent with the relative intensity noise of the multi-longitudinal-mode laser-diode probe. A lower-noise probe would therefore enable a given SNR at lower pump power.

Increasing pump power also produces a progressive blue shift of the water absorption band (Fig. 5d). This shift is consistent with pump-induced local heating, which redistributes water molecules toward weaker hydrogen-bonding environments, as reported in the temperature-dependent water spectra of Ref. [30]. Following the framework developed by Kawai and Shirakashi [31], we decomposed the $H_2O$ spectra acquired at 2.4 and 20.8 mW into three components assigned to weakly ($S_0$), intermediately ($S_1$), and strongly hydrogen-bonded ($S_2$) water (Figs. 5e, f; fitting details in the Methods section). At higher pump power, the fractional fitted area of the weakly bonded component increases, while those of the intermediate and strongly bonded components decrease. Within this three-component model, this redistribution supports the thermal interpretation of the observed blue shift. With temperature calibration, the band shift could provide a local water-temperature readout, while changes in the fitted component areas could be used to examine hydrogen-bond redistribution within the three-component model [31]. Increasing pump power thus improves SNR while also perturbing the sample spectrum. For temperature-sensitive samples, the gain in SNR must therefore be balanced against spectral fidelity to the unperturbed molecular state.

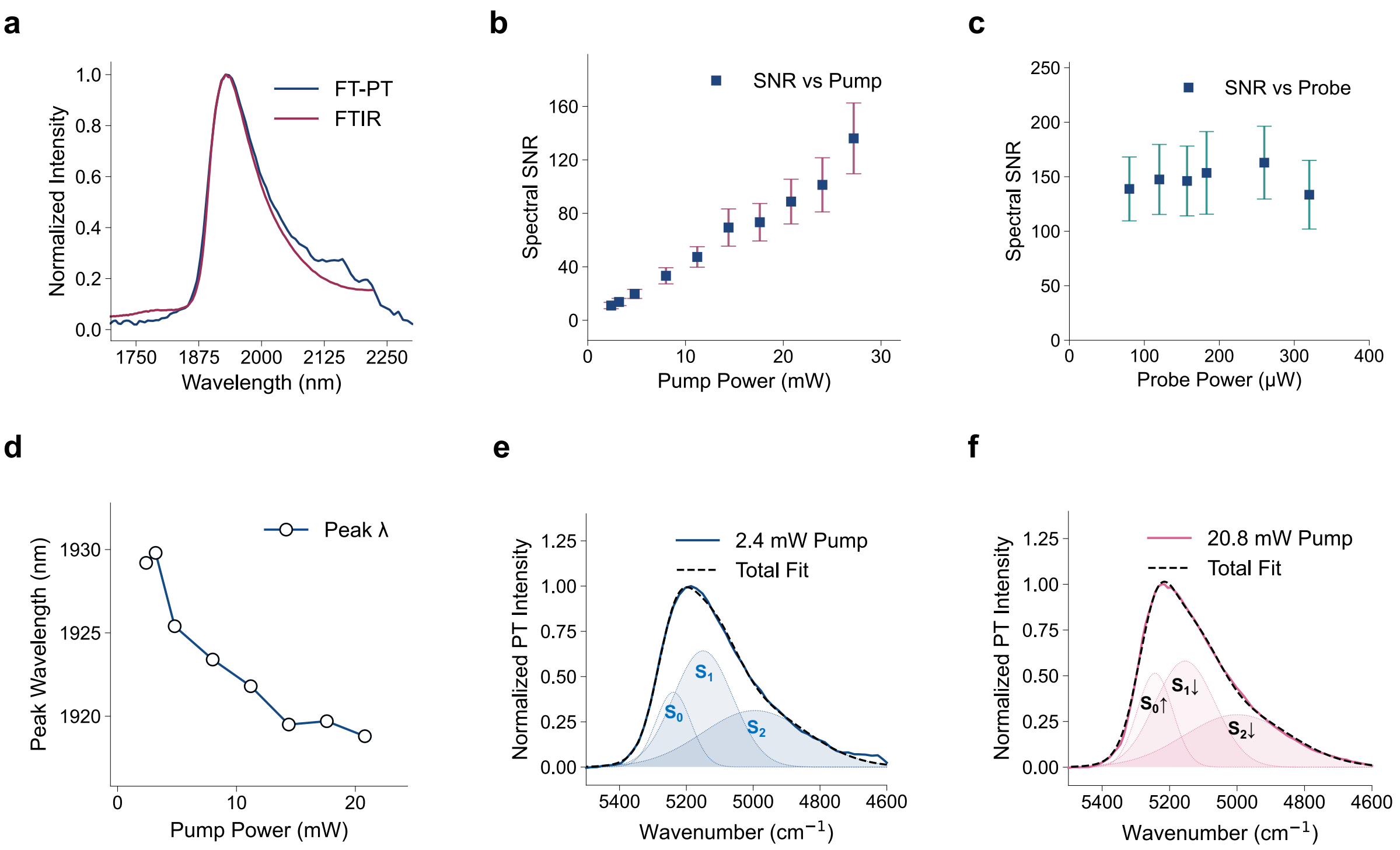

**Fig. 5 Spectral fidelity, SNR, and pump-induced spectral perturbations in liquid water.** (a) Normalized PT spectrum of pure water acquired at 2.4 mW pump power, compared with the reference spectrum measured using a commercial FTIR spectrometer. (b, c) Spectral SNR as a function of pump power and probe power, respectively. (d) Plateau-centroid peak wavelength of the water combination band as a function of pump power, showing an overall blue shift. The centroid is calculated from the upper 4% of the smoothed band. (e, f) Three-component decomposition of the normalized water spectra at pump powers of 2.4 and 20.8 mW, respectively; shaded areas show the individual components, and dashed black curves show the total fits.

## Discussion

The value of SC-driven FT-PT microspectroscopy lies in combining broad spectral coverage with localized optical readout in a table-top implementation. SC excitation supplies spatially coherent light across widely separated overtone and combination bands, while interferometric encoding acquires those bands without wavelength tuning. Microjoule-level pulse energy supports weak-band detection, while nanosecond excitation deposits that energy well within the microsecond thermal-relaxation time. A repetition rate of tens of kilohertz allows substantial cooling between pulses, and the fiber architecture brings these properties together in a compact source. Multiplexed PT spectroscopy and microscopy have been established with frequency combs and synchrotron radiation [20–24]. Our contribution is a rapid-scan implementation for broadband measurements of condensed-phase samples, in which the focused SC pump produces a PT response large enough for direct interferogram acquisition without a lock-in amplifier.

Direct acquisition links measurement speed to spectral fidelity. In a continuously scanned interferometer, optical wavenumber is mapped to temporal modulation frequency. A lock-in output filter with insufficient bandwidth would therefore attenuate and phase-shift different spectral components unequally, altering relative band amplitudes and profiles. Avoiding this additional filter removes that source of spectral weighting and the point-by-point settling required in the reported synchrotron step-scan implementation [24]. The low-power water spectrum agrees with the FTIR reference (Fig. 5a), supporting the spectral fidelity achieved with the present detection bandwidth. The approximately 1-s acquisition at 10 $cm^{-1}$ resolution resolves broad overtone and combination bands. Finer resolution requires a longer OPD scan; faster scanning over the same OPD range provides fewer pump pulses per spectrum and can reduce SNR. Coherent averaging improves weaker features at the expense of temporal resolution.

The single-particle measurement demonstrates the value of combining broadband excitation with spatially selective PT detection. Dark-field readout selects bead-scattered probe light, and several characteristic PS bands are already visible before solvent subtraction. Subtracting the $D_2O$ contribution separates these particle features from the overlapping solvent response. The result extends the platform's utility beyond homogeneous liquids and provides a basis for spectroscopic characterization of individual microscopic objects within heterogeneous samples. The water measurements highlight the ability to detect changes in the local molecular environment through band positions and profiles. The power-dependent water band suggests a route to local thermometry and hydrogen-bond analysis after calibration. The measured power dependences are consistent with multiplicative probe noise, suggesting that a lower-noise probe could achieve a given SNR with less pump power. Improving probe stability therefore offers a route to more sensitive measurements while reducing thermal perturbation and better preserving the sample's molecular state.

Further development could broaden the range of chemical information accessible with this platform. In particular, extending the current near-infrared spectral window into the mid-infrared using supercontinuum sources based on fluoride [32] or chalcogenide [33] fibers could enable access from overtone to fundamental vibrational bands, providing richer and more chemically specific molecular signatures. Adding spatial scanning would allow the demonstrated single-particle spectroscopy to develop into hyperspectral mapping, combining localized detection with broad spectral information at each position. With characterization of spatial resolution and control of thermal perturbation, this approach could support studies of chemical heterogeneity in microscopic materials and, subsequently, biological specimens.

# Methods

**Supercontinuum source configuration**

A 1550-nm seed laser (Shanghai Connet Laser Technology) with a tunable pulse repetition rate (30 kHz–1 MHz) and pulse duration (0.5–10 ns) was employed. An isolator and a fused pump/signal combiner coupled its output to an approximately 2.5-m double-clad erbium–ytterbium co-doped gain fiber, forward-pumped by a 976-nm multimode laser diode. The amplified output was launched into approximately 20 m of standard single-mode silica fiber (SMF-28). At a diode pump power of 9 W, the collimated SC output exceeded 0.6 W, with more than 300 mW retained after long-pass filtering. The combiner, gain fiber, SMF-28 fiber, and pump diode were mounted on a baseplate measuring less than 31 cm × 31 cm × 2 cm, excluding the seed laser and external power supplies.

**FT-PT microspectroscopy experimental setup and data acquisition**

The FT-PT microspectroscopy system employed the SC source described above as the pump. A fiber-coupled 515-nm laser diode (LP515-SF3, Thorlabs) was used as the probe, while a He–Ne laser (HNL100LB, Thorlabs) provided the OPD reference. The pump was focused onto the sample using a 100× reflective objective (891-0003, numerical aperture (NA) = 0.8, PIKE Technologies), and the transmitted probe beam was collected using a 60× objective lens (LUCPlanFLN, NA = 0.7, Olympus). PD1 and PD3 were silicon photodetectors (PDA36A-EC, Thorlabs), whereas PD2 was an extended InGaAs photodetector (PDA10D2, Thorlabs). The PT and OPD-reference signals were simultaneously digitized at 100 kS/s using an ADALM2000 digitizer (Analog Devices).

**Photothermal decay analysis**

Using route 1 in Fig. 2a, we analyzed a 0.6-s segment of the DC-coupled PT recording containing 36001 pulse decays. To examine pulse-to-pulse kinetics without waveform averaging, each decay was individually fitted over a 12 μs window (120 points, sampled at 10 MS/s), beginning 0.5 μs after the peak and ending at least 0.8 μs before the next pump pulse. The relaxation traces were modeled by an effective single-exponential decay, $V(t) = C + A \exp(-t/\tau)$, where A, C, and $\tau$ represent the signal amplitude, baseline offset, and effective 1/e decay time, respectively. All the fits satisfied $R^2 \geq 0.90$. For comparison with heat diffusion models, the room-temperature thermal diffusivities of water and DMSO from Ref. [26] were linearly averaged as a first-order approximation for the mixture.

**Spectral reconstruction and SNR analysis**

For spectral reconstruction, the recorded PT interferogram was resampled at the zero crossings of the He–Ne reference ($\lambda$ = 632.816 nm), providing a uniform OPD interval of 316.408 nm. Individual interferograms were corrected for scan-direction reversal and OPD offset before coherent averaging. A triangular apodization window was applied prior to the fast Fourier transform (FFT), with the spectral axis calibrated using the He–Ne-defined OPD spacing. The spectral SNR was defined as the maximum spectral amplitude across 1600–2400 nm divided by the baseline SD over 2700–2900 nm, corresponding to an electrical frequency window of 1.38–1.48 kHz that lies adjacent to the signal band (1.67–2.50 kHz) to estimate nearby broadband noise outside the SC excitation band.

**Spectral repeatability analysis**

For Fig. 3d–f, 98 individual approximately 1-s acquisitions were analyzed before excitation-profile correction. Each band was fitted in wavenumber with a Gaussian peak plus a local linear baseline. Fitting windows corresponding to 1710–1760, 1840–2050, and 2220–2300 nm were used for the bands near 1730, 1936, and 2270 nm, respectively. The center bounds were 1720–1745, 1910–1960, and 2250–2290 nm to prevent fitting an adjacent peak. FWHM was calculated from the fitted Gaussian width. Peak amplitudes were extracted from individual spectra without per-spectrum intensity normalization, so that acquisition-to-acquisition amplitude fluctuations were preserved. Failed fits and fits with centers or widths at their bounds were excluded. The valid fit counts were 92, 98, and 98 for the three bands, respectively. Means and sample SDs were calculated from the valid fits.

**Single-particle measurements and solvent-background subtraction**

6 μm PS microspheres were drop-cast onto a coverslip, and ethanol was added to assist spreading. After air-drying, the microspheres were physically immobilized on the coverslip. $D_2O$ was then placed on the sample, and a second coverslip was put on top to form a liquid sandwich. The edges were sealed with nail polish to prevent evaporation. For Fig. 4d, 30 aligned interferograms were coherently averaged before Fourier transformation and excitation-profile correction. A pump power of 28 mW was used for the PS bead PT spectrum measurement. The $D_2O$ background was estimated from the tabulated extinction coefficient reported by Wang et al. [34], converted to an absorption coefficient using $\alpha = 4\pi k/\lambda$, where k is the imaginary part of the complex refractive index. A local baseline was estimated from ±0.012 μm windows around 1.80 and 2.20 μm. The $D_2O$ curve was scaled by least squares to the baseline-referenced signal over 1.96–2.05 μm and subtracted from the corrected spectrum.

**Power-dependent water measurements**

Pump power was adjusted by neutral-density attenuation downstream of SC generation while the source operating conditions were held constant. Pump and probe powers were varied separately for the SNR measurements, using the spectral SNR definition above. The normalized FT-PT microspectroscopy and commercial FTIR spectra in Fig. 5a were compared by band position and shape. For Fig. 5d, the spectrum over 1750–2250 nm was smoothed using a second-order Savitzky–Golay filter with a 21-point window. The reported peak wavelength was the intensity-weighted centroid of points at or above 96% of the smoothed band maximum. This reduces sensitivity to noise on the broad peak plateau without assuming sub-nanometer absolute accuracy.

**Three-component decomposition of water spectra**

Near-infrared photothermal spectra spanning 4500–5500 $cm^{-1}$ were analyzed using the three-component framework of Ref. [31]. To distinguish changes in component contributions from a shift of the overall band, spectra acquired at 2.4 and 20.8 mW were jointly fitted by nonlinear least-squares regression, minimizing the sum of squared residuals across both datasets simultaneously. To reduce parameter covariance, the Gaussian widths and relative inter-peak spacings were shared across both pump powers, with one common shift between the spectra, while the three component amplitudes were independently optimized. Identical parameter bounds were applied to both spectra; no constraint imposed a direction of change in the component areas. Initial peak center values of 5229, 5135, and 4986 $cm^{-1}$ were derived from established pure-water reference spectra at 25 °C [35]; these were starting values rather than fixed fitted centers. Before fitting, a linear baseline estimated from 5450–5500 and 4565–4600 $cm^{-1}$ was subtracted, and each spectrum was scaled to its baseline-corrected maximum. The fractional contribution of each component was calculated as its integrated Gaussian area divided by the summed areas of all three components.

**Fundings**

This work was supported by JSPS KAKENHI (25H01386, T.I.), JST FOREST Program (JPMJFR236C, T.I.), and RIKEN TRIP initiative (HIKARI-COOL Tokyo) (T.I.). S.W. was supported by the China scholarship council (202406070117).

**Acknowledgement**

We thank Dr. Keiichiro Toda for his valuable advice on photothermal measurements.

**Data and code availability**

The data and code presented in the manuscript are available from the corresponding author upon reasonable request.

**Author contributions**

T.I. conceived and led the project, and provided overall scientific supervision. J.L. and S.W. designed the supercontinuum source. S.W. performed the experiments and analyzed the data, with contributions to the experiments from Z.X. Z.X. and K.H. provided technical advice and suggestions throughout the project. S.W., J.L., and T.I. wrote and revised the manuscript, with input from all authors.

**Competing interests**

The authors declare no competing interests.